\documentclass[11pt,epsfig]{article}
\usepackage{epsfig}
\title{\bf Five Years of Magellanic Clouds Research:\\ \medskip
A Newsletter Editors' Perspective}

\author{E.~K.~Grebel$^1$, Y.-H.~Chu$^2$, J.~S.~Gallagher$^3$, D.~Harbeck$^1$\\
\vspace{0.1cm}\\
\normalsize $^1$ Max-Planck-Institut f\"ur Astronomie, K\"onigstuhl 17, D-69117
                 Heidelberg, Germany\\
\normalsize $^2$ Department of Astronomy, University of Illinois, 1002 W.\ 
                 Green St., Urbana, IL 61801, USA\\
\normalsize $^3$ Department of Astronomy, University of Wisconsin, 475 N.\ 
                 Charter St., Madison, WI 53706, USA\\
}
\date{}
\begin{document}
\maketitle
\def\bull{\vrule height .9ex width .8ex depth -.1ex}
\makeatletter
\def\ps@plain{\let\@mkboth\gobbletwo
\def\@oddhead{}\def\@oddfoot{\hfil\tiny
``Dwarf Galaxies and their Environment'';
International Conference in Bad Honnef, Germany, 23-27 January 2001}%
\def\@evenhead{}\let\@evenfoot\@oddfoot}
\makeatother

%%  if your contribution is short, you may, if the title is clear enough, 
%%  skip the abstract.....
\begin{abstract}\noindent
 We analyze the topical and demographic evolution of Magellanic Clouds
 research over the past five years based on submissions of abstracts of
 refereed papers to the electronic Magellanic Clouds Newsletter
 ({\tt http://www.astro.uiuc.edu/projects/mcnews/MCNews.html}).
\end{abstract}

\section{Introduction}

In late 1995 the Magellanic Clouds Newsletter (MCNews) was founded by 
You-Hua Chu's Magellanic Clouds Working Group at the University of 
Illinois in Urbana-Champaign (UIUC).  Two years earlier a joint
``Graduiertenkolleg'' (graduate school) for Magellanic Clouds research
had been created by the Universities of Bonn and Bochum.
Several members of the Graduiertenkolleg went to UIUC as exchange
visitors, and the University of Bonn became the European mirror site
for MCNews.  The Graduiertenkolleg is now reaching the end of its
funding period, and we use this opportunity to analyze the topical 
and demographical evolution of Magellanic Clouds research worldwide 
over the past five years as reflected in MCNews.

\section{The Magellanic Clouds Newsletter}

MCNews covers all areas of Magellanic Clouds research and publishes 
abstracts of submitted and accepted
refereed papers, PhD theses, conference proceedings, and job and conference
announcements.  Its editors are Eva Grebel
(MPIA) and You-Hua Chu (UIUC).  
Since September 1997 MCNews appears monthly and is currently sent
out electronically in \LaTeX\ format
to $\sim 440$ subscribers in 31 countries. 
Fifty five issues have appeared as of May 2001, comprising
a total of 532 abstracts of refereed papers (an estimated 80\% of 
the refereed publications in this area), so we estimate that we are 
reaching $\sim 80$ \% of the researchers active in this field.

\section{Demographics of Magellanic Clouds research}

Assuming that MCNews subscribers
are approximately representative of Magellanic Clouds researchers, then
Europe (42\%) and North America (37\%) have the highest concentration of
Magellanic Clouds researchers.  For the top eight countries the
following fractional distribution of subscribers results:
USA: 32\%, Germany: 16\%, Australia: 8\%, France: 7\%, UK: 6\%, Italy:
5\%, Chile: 4\%, Canada: 3\%.  Between 8\% (Asia) and 24\% (South
America) are women (i.e., 18\% on average).

\begin{figure}[ht]
%\begin{flushleft}
\epsfig{file=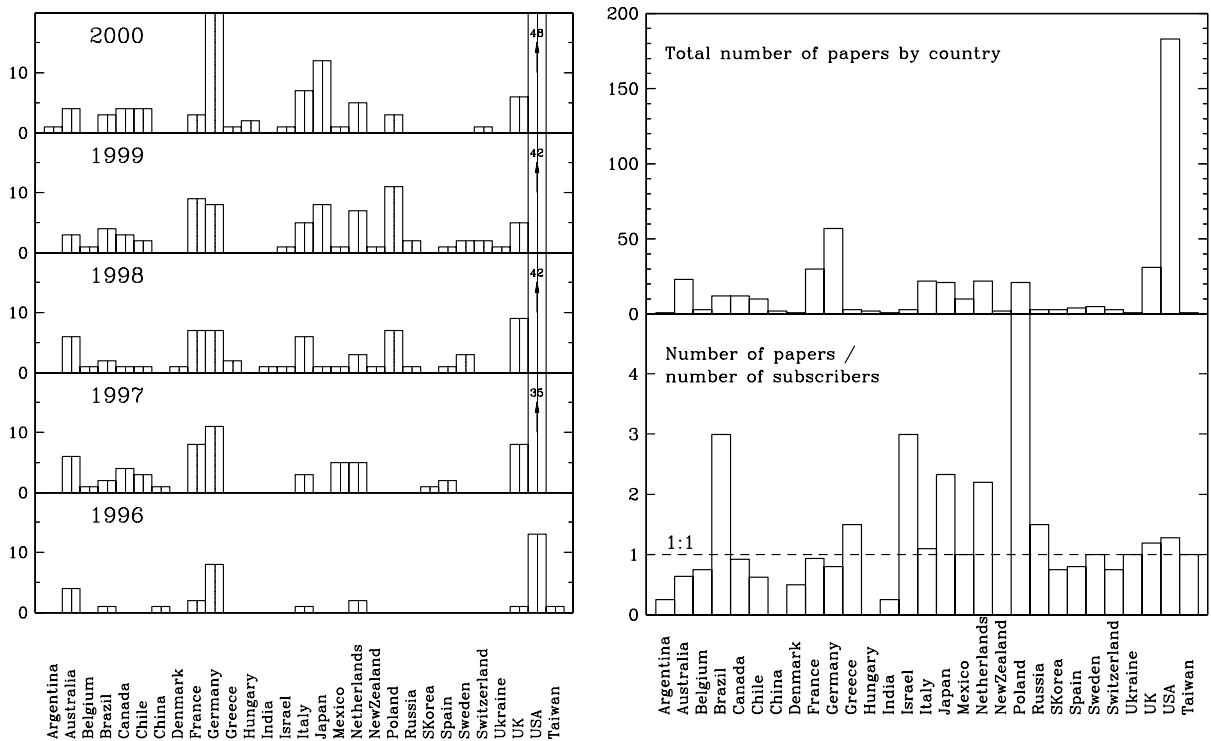, width=17.0cm} 
\vskip-1.5cm
\caption{{\it Left panel:\, }  Number of refereed papers listed in MCNews
per country and year.  The low overall count in 1996 is likely
too low since MCNews was still in its initial phase then.  {\it Upper right
panel:\, } Total number of refereed papers 
per country from 1996 -- 2000.  {\it Lower right panel:\, } 
``Productivity'' measured by dividing the number of refereed papers by the
number of subscribers per country.  Countries with few
subscribers but comparatively large numbers of papers lie well above the
line marking an average of one paper per subscriber. } 
%\end{flushleft}
\end{figure}

\section{Scientific productivity}

The number of refereed papers corresponds roughly to the number of
Magellanic Clouds researchers per country.  The largest total number of
publications is found in the US, followed by Germany, France, and the UK.
We use the affiliation of the first author to assign a paper
to a specific country without considering the rest of the author team.
Some of the annual variations are directly correlated with special campaigns
or instruments.  For example, the 
pronounced increase in Japanese refereed papers
on the Magellanic Clouds is based largely on data from the 
X-ray satellite ASCA.  The activity
peaks in Poland are almost exclusively due to the 
OGLE microlensing experiment.

Dividing the total number of papers per country by the number of subscribers
per country gives an approximate measure of activity per researcher
in the said country.  Countries with few subscribers but very active
research groups stand out in this normalization.  Poland
takes a pronounced lead due to its very productive OGLE group and
small number of subscribers.  Other
countries where researchers have a per capita rate of refereed papers that
clearly exceeds 1 are Brazil (especially star clusters), Israel (particularly
theory), Japan (X-ray binaries), and the Netherlands.  Countries
with a large total number of refereed papers (e.g., USA, Germany) are
close to one paper per subscriber, illustrating that not all of their
subscribers actively work on the Magellanic Clouds.  Further, not
every person who submits an abstract to MCNews is a subscriber,
and the number of papers per subscriber ranges from 0 to $> 20$.

\newpage

\begin{figure}[ht]
%\begin{flushleft}
\epsfig{file=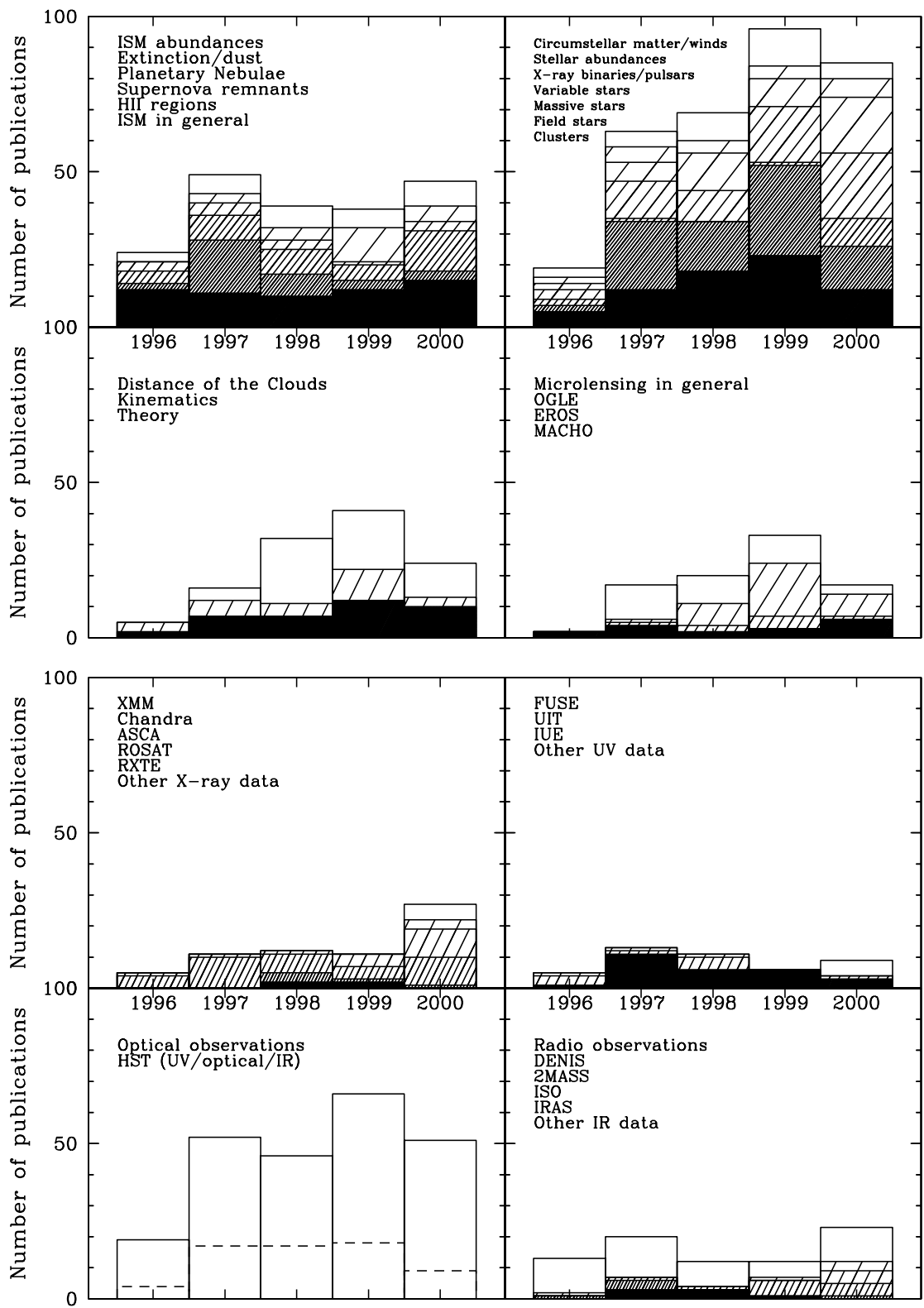, width=16.0cm}
\vskip-1.5cm
\caption{{\it Upper panel:\, } Topics in Magellanic Clouds research by 
year and total number.  The numbers for 1996 are probably incomplete.
The sequence of subdivisions of the histogram bars corresponds to the
sequence in the legend of each plot (white: top item; black: bottom item;
hashed with increasing density: items in between).
{\it Lower panel:\, } Magellanic Clouds research by wavelength/satellite
per year and total number.
}
%\end{flushleft}
\end{figure}

\clearpage

\section{Main areas of Magellanic Clouds research}

Stellar populations -- individual stars such as massive stars, AGB
stars, variables, binaries; field populations, clusters and associations --
account for the largest number of refereed publications
on the Magellanic Clouds.  The fractions of the main research areas 
are as follows: 
Stellar populations:  45\%,  interstellar medium: 27\%, 
microlensing:         12\%, theory:               7\%,  
distance:      5\%, dynamics/kinematics:   3\%. 
Many studies belong to several of these research
areas, each of which was credited in these cases.  Interestingly, theoretical
studies account only for a small fraction of the total.
Theoretical research concentrates mainly on evolutionary models, both for
individual stars as well as for stellar synthesis and chemical evolution.
Many Magellanic Clouds studies are based partially or wholly on optical
data (fractional distribution:  Optical:  53\%, X-ray:    16\%,
UV:       12\%, radio/sub-mm:    11\%, infrared:        8\%).

\section{Milestones of Magellanic Clouds Research}

The past five years have seen significant progress in many areas of 
Magellanic Clouds research, with large-scale surveys playing an 
important role.  To highlight just a few, we recall the 
impact of the Australian H\,{\sc i} synthesis maps of the Clouds, which 
revealed a complex, fractal ISM full of shells, holes, and fragments. 
The HIPASS multi-beam survey detected the leading arm of the Magellanic stream.
The NANTEN CO survey showed the distribution of molecular clouds
at unprecedentedly high resolution and correlated their location 
with the age of star-forming regions.  A growing body of 
high-resolution infrared data (e.g., NICMOS, VLT) is
resolving pillars, Bok globules, and 
pre-main sequence stars in the Clouds.  High-resolution studies of the stellar
IMF in clusters (mostly with HST) show evidence for
Salpeter-like slopes, but also clear indications of mass segregation.
Infrared and optical
imaging surveys (e.g., DENIS, 2MASS, the Magellanic Clouds Photometric 
Survey) unveiled the structure of the Magellanic Clouds as traced by their 
various stellar components and led to comprehensive point source
catalogs of special stellar types such as AGB stars and carbon stars.  
Searches for carbon stars at the periphery of the Clouds, and kinematic
studies resulted in extended rotation curves and the identification of
kinematic subcomponents.  Many studies concentrated on red clump stars
as distance and evolutionary indicators.  The microlensing surveys
(esp.\ EROS, MACHO, OGLE) yielded a wealth of information on variable
stars in the Clouds and contributed to the still unsolved
question of the distance to the Clouds.  Eclipsing binaries are emerging
as one way to solve this question independent of stellar evolutionary
assumptions.  
%Hipparcos data improved our knowledge of the proper
%motion of the Clouds. 
X-ray surveys (e.g., ROSAT, ASCA) are helping to complete
the census of supernova remnants, X-ray binaries, as well as revealing the
large-scale distribution of hot gas.
Deep HST observations (WFPC2)
made possible accurate relative age determinations for the oldest
globular clusters in the Magellanic Clouds, and the derivation of detailed 
star formation histories for their field populations.
OGLE and other imaging surveys led to improved star cluster
catalogs and cluster age-dating. 
Abundances of field stars and clusters were 
derived photometrically (Str\"omgren, Washington) and through increased
spectroscopic samples (Ca\,{\sc ii} triplet), revealing a complex
age-metallicity relation.  High-resolution stellar
abundance measurements are also becoming more available; an
area that will likely expand with the advent of the new large southern
telescopes.  The IUE archive, GHRS, and STIS advanced our knowledge of the 
UV and wind properties of massive stars.
GHRS, STIS, ORFEUS, and FUSE increased our knowledge of
interstellar abundances in the Clouds and led to the detection of H$_2$.
These studies are complemented by radio and sub-millimeter measurements
in the Clouds and in the Magellanic bridge.  Surprisingly, stellar and
gaseous abundances in the bridge turn out to be lower than those of 
the young field populations in both the LMC and SMC.

We look forward to the exciting results that the coming years and the
new large telescopes and satellite missions will bring.

\end{document}